\documentstyle[11pt,roy,twoside,epsf]{article}
\def\edcomment#1{\iffalse\marginpar{\raggedright\sl#1\/}\else\relax\fi}
\reversemarginpar

\begin{document}
\title{Parsec-Scale Jets and Tori in Seyfert Galaxies}
 \author{A.L. Roy$^1$, J.M. Wrobel$^2$, A.S. Wilson$^3$, J.S. Ulvestad$^2$, 
R.P. Norris$^4$, C.G. Mundell$^5$, T.P. Krichbaum$^1$, H. Falcke$^1$, 
\& E.J.M. Colbert$^6$
}
\affil{$^1$MPIfR, Bonn, Germany, aroy@mpifr-bonn.mpg.de}
\affil{$^2$NRAO, PO Box O, Socorro, NM 87801, USA}
\affil{$^3$University of Maryland, College Park, MD 20742, USA}
\affil{$^4$ATNF, PO Box 76, Epping NSW 1710, Australia}
\affil{$^5$ARI, Liverpool John Moores University, Birkenhead, UK}
\affil{$^6$JHU, Dept Physics and Astronomy, Baltimore, MD 21218, USA}

\begin{abstract}

What causes the dichotomy between very powerful and very weak radio
emission from AGNs?  Perhaps the engines are the
same but the jets get disrupted by dense ISM in radio-quiet objects,
or else the engines are intrinsically different with jet power scaling
with, say, black hole spin.  To distinguish, one can look for
interaction between the jets and the NLR and measure the jet speed
close to the core using VLBI, before environmental effects become important.
We find that in radio-quiet AGN, the jets appear slower and have a
greater tendency to bend, and that one-sidedness and flat-spectrum
cores are probably due to obscuration.

%Active galaxies tend to be either very powerful or very weak radio sources.
%Although we have known this for 30 years we still do not understand the
%underlying cause of the dichotomy.  Perhaps the engine is the same in
%both systems and the jets get disrupted by dense interstellar material
%in radio-quiet objects, or else the difference is intrinsic with jet
%power scaling with, say, black hole spin or mass.  To
%distinguish, one can look for signs of interaction between the jets and
%the narrow-line region, and measure the jet speed close to the core
%using VLBI, before environmental effects become important.
%We report results of such studies.

\end{abstract}

\section{Four Differences between Low-Power and High-Power Jets}

1) In our VLBA observations of Seyfert galaxies at frequencies between
1.6~GHz and 15 GHz we have found sub-relativistic jet component speeds
(typically $\leq 0.25\,c$) in several Seyfert galaxies, such as in Mrk
348 at a distance of 0.5 pc from the core, in Mrk~231 at 1.0 pc
(Ulvestad et al. 1999), in NGC 5506 at 3.4~pc (Roy et al. in prep),
and in NGC~1068 at 21~pc (Fig 1; Roy et al. in prep).  In contrast, in
more powerful radio sources such as QSOs, the jet speeds have usually
proven to be relativistic.  The difference could signify either an
intrinsic difference between the engines or a deceleration of the flow
in Seyfert galaxies in the BLR gas on scales $< 0.5$ pc.  (Note though
that if the Seyfert components are shocks, then their speeds
would be different from the jet flow speed.)

\begin{figure}
\plotone{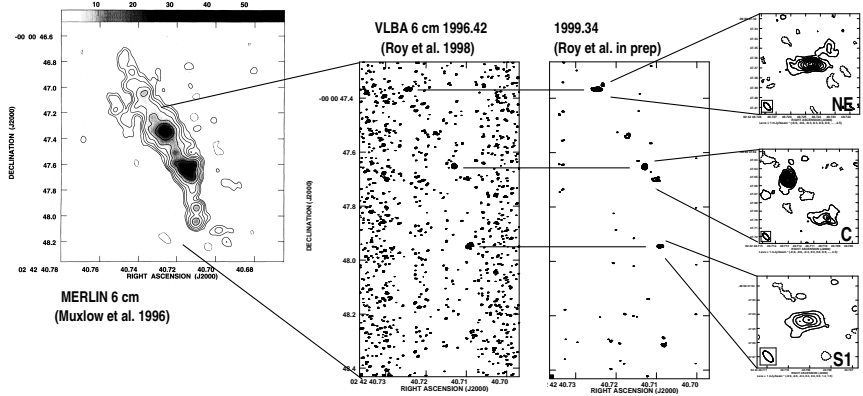}
\caption{Centre and Right: VLBA images of NGC 1068 at 6 cm and resolution 
of (6 $\times$ 4) mas for two epochs, showing no motion ($v < 0.075$~$c$ for 
$H_{0} = 75$ km s$^{-1}$ Mpc$^{-1}$).  Left: MERLIN image at 60 mas
for perspective.}
\end{figure}

2) We found one-sided jets in VLBA images of the Seyfert galaxies Mrk~348,
Mrk 231, and NGC~5506.  In these galaxies the jets are low power
and slow and probably lie across the line of sight (NGC~5506 and Mrk
348 host water masers, indicating edge-on systems, and in Mrk 231 the jet
is inclined at $45^\circ$ if it is perpendicular to the 100 pc scale
H\,I disc; Carilli et al. 1998).  In powerful radio jets the
one-sidedness is usually explained with Doppler boosting, but in these
Seyfert galaxies, boosting should
be small, which leaves free-free absorption of the counterjet by a
screen (perhaps the torus) as a more likely explanation.

3) Flat or absorbed spectra were found in some of the parsec-scale
radio components in Mrk 348, Mrk 231, NGC~2639, and NGC~5506 and we
found that these could be caused by free-free absorption by a
foreground screen.  Such absorption would be expected if the X-ray
absorbing column seen in, say, Mrk~348 (10$^{27}$ m$^{-2}$), occurred
in a slab 0.1 pc thick like the expected inner torus edge; and is enough
to extinguish our view
of a counter-jet.  However, note that the $T_{\rm b}$ values we
measured are lower limits of $\simeq 10^7$ K, and so $T_{\rm b}$ could
in principal be as high as $10^{10}$ K and then the spectra could be
due to synchrotron self absorption (SSA).  In contrast, flat-spectrum
cores seen in high-power jets are rarely attributed to free-free
absorption and are usually explained as SSA.

4) The jet in NGC 5506 bends through $90^\circ$ at 3.4 pc to align
with the large-scale outflow, and in Mrk 231 the jet base is
misaligned by $65^\circ$ with the 40 pc-scale lobes.  The bends are
probably intrinsic because the jet inclination is large, whereas in
powerful jets such bends are usually interpreted as projection
effects.  The bends in Seyferts might be a sign of interaction between the
jet and the NLR gas on parsec scales, and indeed by momentum arguments, 
low-power jets are more easily bent than high-power jets.

\end{document}